\documentclass{article}

\usepackage{arxiv}

\usepackage[utf8]{inputenc} 
\usepackage[T1]{fontenc}    
\usepackage[spanish]{babel}
\usepackage{hyperref}       
\usepackage{url}            
\usepackage{booktabs}       
\usepackage{amsfonts}       
\usepackage{nicefrac}       
\usepackage{microtype}      
\usepackage{graphicx}
\usepackage{tabularx, booktabs}
\usepackage{natbib}
\usepackage{doi}

\title{CONSTELACIONES MAYAS: VISUALIZACIÓN E INTERPRETACIÓN UTILIZANDO HERRAMIENTAS INFORMÁTICAS}

\date{November 30,2020}	

\author{ \href{https://orcid.org/0000-0001-6956-0987}{\includegraphics[scale=0.06]{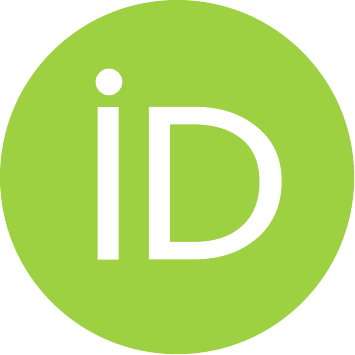}\hspace{1mm}Eduardo Enrique Rodas-Quito}\\
	Department of Archaeoastronomy and Cultural Astronomy\\
	Space Sciences Faculty\\
	Universidad Nacional Autonoma de Honduras \\
	\texttt{eduardo.rodas@unah.edu.hn} \\
	\And
    \href{https://orcid.org/0000-0001-8936-7236}{\includegraphics[scale=0.06]{orcid.pdf}\hspace{1mm}Javier Mejuto González}\\
	Department of Archaeoastronomy and Cultural Astronomy\\
	Space Sciences Faculty\\
	Universidad Nacional Autonoma de Honduras \\
	\texttt{javier.mejuto@unah.edu.hn} \\
}

\renewcommand{\shorttitle}{CONSTELACIONES MAYAS : VISUALIZACIÓN E INTERPRETACIÓN}

\begin{document}
\maketitle

\renewcommand{\abstractname}{RESUMEN}
\begin{abstract}
\noindent Se ha debatido por mucho tiempo acerca de la existencia de constelaciones dentro de la cultura maya. Las evidencias que apoyan esta hipótesis provienen de diversas fuentes: pinturas en murales, estelas y códices, entre otros.  En este trabajo, hemos recuperado y compilado información acerca de las constelaciones mayas y los nombres de algunos cuerpos astronómicos seleccionados.  Se explica el proceso llevado a cabo para obtener dicha información y recopilarla en un soporte digital.  El resultado fue incorporado en el programa de cómputo astronómico Stellarium, para enriquecer los módulos de arqueoastronomía y leyendas estelares, en donde se han puesto a disposición de los usuarios las cons-telaciones de diversas culturas, simulando su apariencia en el cielo nocturno.

\vspace{0.3cm}
\noindent \textbf{Palabras clave:} Stellarium, constelaciones mayas, educación, vinculación con la sociedad.
\vspace*{0.1cm}
\end{abstract}

\renewcommand{\abstractname}{ABSTRACT}
\begin{abstract}
\noindent There has been a long standing debate about the existence of maya constellations, but there are several sources of evidences that favour this hypothesis: mural paintings, stelae and codex.  In this work, we have recovered and compiled information regarding mayan constellations and the mayan names for selected celestial bodies.  We explain the process that was carried out to obtain all the information available and compile it under the same digital support.  The result was incorporated into the Stellarium astronomical software to enrich the modules of archaeoastronomy and stellar legend, where constellations from diverse cultures are made available to users of the software by simulating their appearance in the night sky.

\vspace{0.3cm}
\end{abstract}

\vspace{2cm}

\section{\textsl{Introducción}}

Varias civilizaciones de la antigüedad imaginaron que las estrellas visibles en el cielo formaban patrones semejando figuras que asociaban a sus diversas creencias, ya fueran mitológicas, religiosas o históricas, a las cuales conocemos como constelaciones. Así tenemos a los antiguos chinos que desarrollaron un sistema de patrones de estrellas en el que proyectaron su sociedad y cultura, diferente del que inventaron los antiguos griegos (\citeauthor{sun2015}, \citeyear{sun2015}: 2052). Los antiguos egipcios creyeron ver en el cielo algunos elementos culturales asociados con sus ritos funerarios (\citeauthor{lull}, \citeyear{lull}: 222).  En Mesopotamia crearon algunas figuras celestes que luego pasaron a nuestra civilización occidental, tal como lo evidencian algunos restos arqueológicos en los que se han identificado varias de las actuales constelaciones (\citeauthor{belmonte1999}, \citeyear{belmonte1999}: 95; \citeauthor{pizzimenti}, \citeyear{pizzimenti}).  Es de esperar, por tanto, que la maya, otra de las grandes civilizaciones de la antigüedad, tuviera también algún tipo de sistema de constelaciones o patrones celestes que pudieran haber utilizado con alguno de los propósitos descritos.

La primera evidencia de que hubo constelaciones en las culturas mesoamericanas nos la proporciona Fray Bernardino de \citeauthor{sahagun1985} (\citeyear{sahagun1985}: 129-130), al mencionar grupos de estrellas a los que los aztecas habían dado un nombre propio: Yoaltecutli, Yacahuitztli, Mamalhuaztli y Citlalxonecuilli.  Otro investigador que sugirió que los mayas también tenían grupos de estrellas con nombres propios fue Alexander von Humboldt (\citeauthor{spinden1916}, \citeyear{spinden1916}: 53), quien en su tiempo fue ampliamente citado al respecto.  A pesar de lo anterior, a principios del siglo XX, todavía no existía un consenso entre los investigadores, cuando algunos de ellos planteaban que las constelaciones estaban representadas en libros mayas (llamados códices), mientras otros debatieron las razones para justificar si los  pueblos originales de centroamérica contaron con un conjunto de constelaciones zodiacales, según reporta \citeauthor{spinden1916} (\citeyear{spinden1916}: 53-54).  Ese mismo autor intenta ubicar la posición de algunas constelaciones zodiacales en el cielo (\citeauthor{spinden1924}, \citeyear{spinden1924}: 55, figura 25), pero no es sino hasta la segunda mitad del siglo XX, coincidiendo con la interpretación de la escritura maya, que empieza a surgir entre los investigadores una convergencia de ideas acerca de la identificación y ubicación en el cielo de las constelaciones mayas.

Las fuentes principales de evidencias para la identificación de las constelaciones mayas son la escritura e iconografía de estelas, fachadas de edificios, monumentos y códices.  Actualmente se dispone de cuatro códices: el Códice de Dresde, que contiene información referente a los ciclos de observación y desaparición del planeta Venus, los ciclos de las lunaciones y su posible relación con eclipses de Sol y de Luna; el Códice de Madrid, que muestra relaciones entre la agricultura y la observación astronómica; el Códice de París, del que se ha presumido por más tiempo que contiene información referente a la observación de constelaciones en el cielo por los mayas y el Códice Maya de México (anteriormente conocido como Códice de Grolier).

\begin{figure}[h]
  \centering
  \includegraphics[width=.39\linewidth]{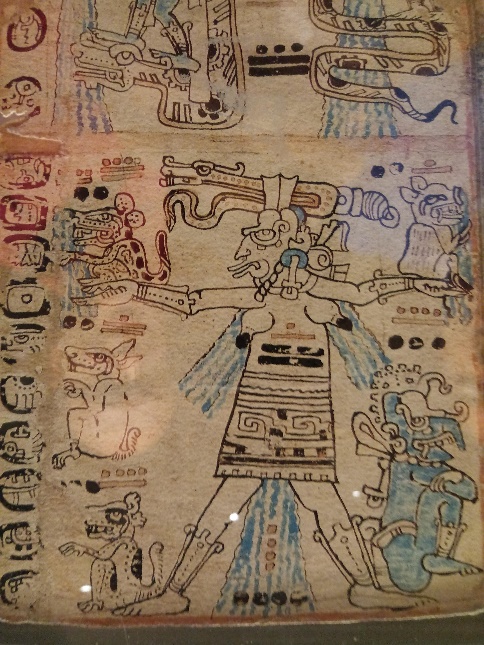}
  \caption{Fotografía de una sección de la réplica del Códice de Madrid en el Museo de América, Madrid, España. (Créditos: los autores).}
  \label{Art004:f2}
\end{figure}

\begin{figure}
    \centering
    \includegraphics[scale=0.8]{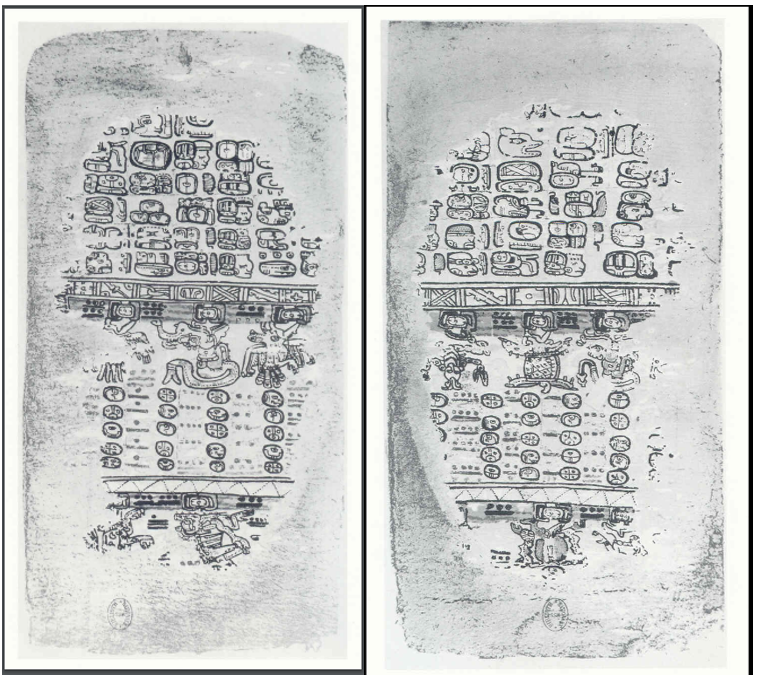}
    \caption{Páginas 23 y 24 del Códice de París. \citep{FAMSIa}}
    \label{Art004:f3}
\end{figure}

Se cree que el Códice de París fue llevado a Europa por los conquistadores como una evidencia de los hallazgos que los descubridores y conquistadores relataron en sus crónicas, pero pronto fue olvidado, al igual que los demás códices conocidos hoy en día.  A pesar de algunos recuentos que aparentemente lo mencionan (en la primera mitad del siglo XIX) \citep{spotak}, no fue sino hasta el año 1859, cuando el orientalista francés León de Rosny lo encontró en la esquina de una chimenea en la Biblioteca Nacional de París, que se supo nuevamente de su existencia (\citeauthor{coe}, \citeyear{coe}: 101).  Como se puede apreciar en la Figura \ref{Art004:f3}, este documento ha sufrido bastantes daños en su estructura, especialmente en los bordes, lo que ha dificultado su lectura e interpretación por parte de los estudiosos del lenguaje maya.  A pesar de esto, se ha determinado por parte de varios autores que el Códice de París describe en las páginas 23 y 24 a las constelaciones de un zodíaco maya, así como los períodos de tiempo que el Sol tarda en pasar entre ellas (\citeauthor{aveni}, \citeyear{aveni}: 275-280; \citeauthor{freidel}, \citeyear{freidel}: 99-103; \citeauthor{spotak}, \citeyear{spotak}).

Han existido varias interpretaciones tanto de los seres representados en el Códice de Dresde como de las posiciones que deberían ocupar en el cielo, por lo que en la próxima sección se describirán los criterios utilizados para la identificación de criaturas finalmente colocadas en el programa de cómputo, así como los métodos implementados para ubicarlos en el cielo, sobre el fondo estrellado.  En esa misma sección se describe cómo el programa de cómputo Stellarium permite incorporar en su operación a constelaciones de culturas diferentes a la occidental, así como los tipos de datos necesarios para poder llevar a cabo dicha actividad.  Como resultado, se presentan los seres que conforman las constelaciones mayas y una descripción de los mismos, junto con la explicación de su posición final en el cielo, tanto textual como gráficamente.  Finalmente se comentan algunos de los problemas que se enfrentaron y explicaciones alternas a las aquí implementadas, así como las razones por las que no se les tomó finalmente en cuenta para las representaciones implementadas.

\section{\textsl{Metodología}}

Existen diversas interpretaciones de las imágenes presentadas en los códices mayas, en las vasijas encontradas en entierros y en estructuras monumentales.  En el caso particular que motivó este trabajo, es decir, implementar una representación visual de las antiguas constelaciones mayas utilizando un programa de cómputo que simulara el cielo nocturno, se buscó en diversas fuentes para obtener la información pertinente a este aspecto de la cultura maya.  Dada la naturaleza del trabajo, se buscaron fuentes de dos tipos: gráficas, para tomar de ellas el aspecto de las figuras a representar en el cielo y las documentales, para dar el soporte argumentativo al porqué de la decisión de usar determinadas figuras.  Las fuentes gráficas son:

\begin{itemize}
    \item Códice de París
    \item Mural de Bonampak
    \item Imágenes en estelas
\end{itemize}

A continuación, se explica el proceso de selección de imágenes celestes dentro de cada una de estas fuentes.

\subsection{\textsl{Códice de París}}

Este documento contiene imágenes que se interpretan como constelaciones que los mayas idearon para identificar la posición del Sol.  Dichas imágenes se encuentran en las páginas 23 y 24 del códice (Ver Figura \ref{Art004:f3}).  La forma en cómo se leen estas figuras en el códice es de derecha a izquierda. Esto, según \citeauthor{tedlock1999} (\citeyear{tedlock1999}: 44-45), es debido a dos razones: en primer lugar, las figuras están con su cuerpo, o partes de éste, orientado hacia la derecha, en oposición al orden de lectura normal de los textos mayas.  La segunda razón es que este orden de derecha a izquierda se ajusta a la relación entre el comportamiento de los fenómenos en la tabla zodiacal del códice con otras tablas. El Sol, la Luna y los planetas, en su movimiento diario, van de Este a Oeste, conforme al movimiento de rotación de la Tierra.  Sin embargo, al observarlos a lo largo de varios días, su movimiento sideral es de Oeste a Este, es decir, para un observador que lleve el registro de este último movimiento, estos cuerpos astronómicos se mueven de derecha a izquierda si se observa utilizando el punto cardinal Sur como referencia, por tanto, de esa misma forma lo plasmaron en el Códice de París.  Para definir el orden en el que se debería colocar en el cielo a cada figura de constelación, se siguió el método descrito por Linda Schele (\citeauthor{freidel}, \citeyear{freidel}: 101-102). Según dicha investigadora, en el ordenamiento de las figuras como constelaciones, se deben colocar las figuras adyacentes en el códice tomando en cuenta los números escritos entre ellas, que representan la cantidad de días que deben transcurrir para que el Sol haga el recorrido entre las constelaciones a lo largo de la eclíptica.  De esto se deduce que estas figuras de constelaciones, aunque aparecen adyacentes en el códice, no lo están en el cielo, sino que están separadas por algo menos de medio año (si tomamos el número escrito entre las figuras, 168 , como número de días necesarios para que el Sol pase de una figura a otra y los dividimos entre 30 días promedio que tiene cada mes,  equivalen a 168/30 = 5.6 meses).  El punto de partida es la constelación del Escorpión, que en la cultura azteca, también de Mesoamérica, corresponde al mismo grupo de estrellas en el cielo que en varias culturas, tal como lo dice \citeauthor{sahagun1985} (\citeyear{sahagun1985}: 130): “A aquellas estrellas, que en algunas partes llaman el Carro, esta gente las llama Escorpión, porque tienen figura de escorpión o alacrán, y así se llaman en muchas partes del mundo”.  La investigadora Linda Scheele también llega a la conclusión que la constelación del Escorpión de la cultura occidental es la misma que la de los mayas, por medio de la interpretación de la iconografía presente en la estela maya conocida como Hauberg (\citeauthor{freidel}, \citeyear{freidel}: 102-104).

\subsection{\textsl{Mural de Bonampak}}

\begin{figure}
    \centering
    \includegraphics[scale=0.73]{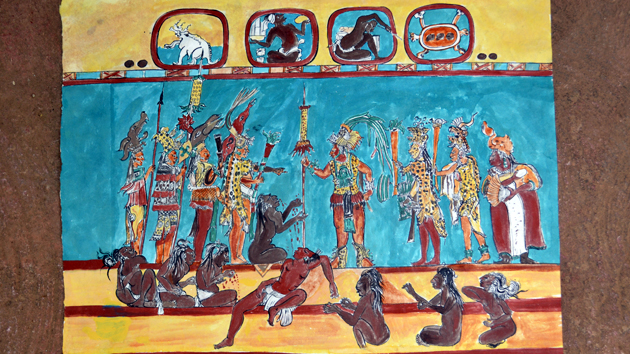}
    \caption{Reproducción del mural en el Cuarto II del Edificio de las Pinturas, en Bonampak, Chiapas, México (Crédito: Patricia Martín Morales).}
    \label{Art004:f4}
\end{figure}

En esta antigua ciudad maya del sureste de México se han encontrado bóvedas decoradas con diversos motivos.  Uno de ellos muestra al gobernante Yahaw Chan Muwan juzgando a cautivos de guerra, bajo un cielo en el que se muestran figuras enmarcadas por cartuchos (ver Figura \ref{Art004:f4}).  Floyd Lansbury propuso que dichas figuras representan a la constelaciones de Géminis y Orión en base al análisis de la fecha en que ocurren los eventos representados en el mural (\citeauthor{mmiller1986} ,\citeyear{mmiller1986}: 30-38).  La tortuga en el extremo derecho del mural se ajusta muy bien con el asterismo occidental del Cinturón de Orión, dada la similitud entre las estrellas visibles sobre el caparazón de la tortuga con el asterismo antes mencionado. Los personajes en los cartuchos centrales representan los planetas Marte y Saturno, todos ellos visibles en los amaneceres de inicios del mes de agosto del 792 d.C.  \citeauthor{galindo2009} (\citeyear{galindo2009}: 26 – 27) y Linda Schele (\citeauthor{freidel}, \citeyear{freidel}: 79 – 81) han aportado más evidencias que apoyan a esta hipótesis.

\subsection{\textsl{Imágenes en estelas y/o monumentos}}
\begin{figure}
    \centering
    \includegraphics[scale=0.85]{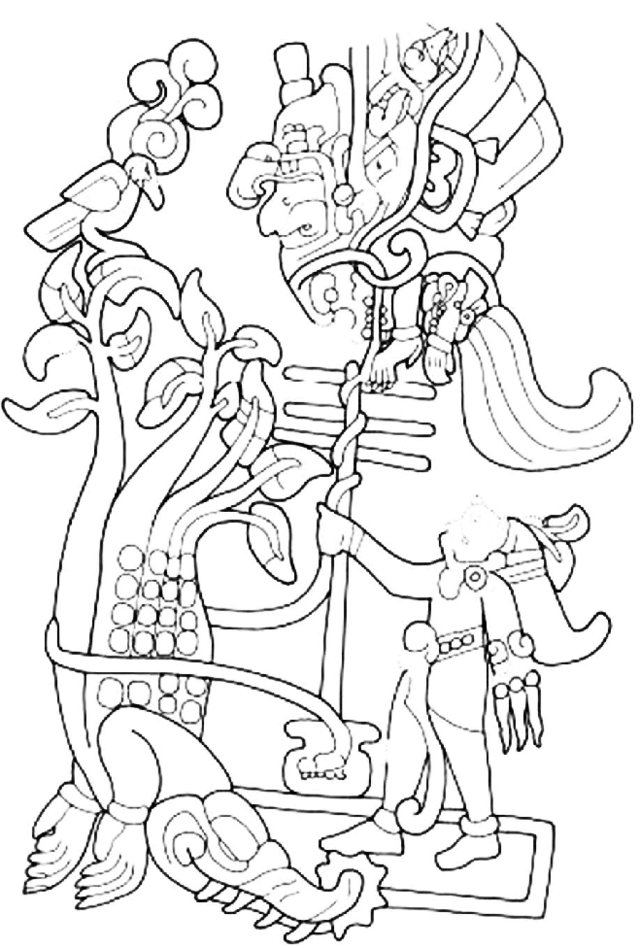}
    \caption{Dibujo de la Estela 25 de Izapa \citep{ochm2010}.}
\label{Art004:f5}
\end{figure}

Sabemos que escenas de la mitología maya se representaron no solo en murales o códices, sino que también en otros soportes. Un ejemplo de esto lo tenemos en la estela 25 del sitio arqueológico de Izapa, México (Figura \ref{Art004:f5}). En ella se observa la representación de un evento que tuvo lugar en el tiempo mítico, antes de la creación según la tradición maya.  Se observa a uno de los gemelos héroes sosteniendo un poste sobre el cual está el dios Itzam-Yeh posando en forma de pájaro.  Del poste cuelga un monstruo con forma combinada de cocodrilo y árbol: el mítico cocodrilo-venado-estelar, o Monstruo Cósmico (\citeauthor{freidel}, \citeyear{freidel}: 88-89), que al ser sacrificado por el dios L del panteón maya, provocó un diluvio con su sangre derramada y con la cual se llevó a cabo la creación del mundo (\citeauthor{garciab2015}, \citeyear{garciab2015}; \citeauthor{velasquezg2006}, \citeyear{velasquezg2006}).

\begin{figure}
    \centering
    \includegraphics[scale=0.67]{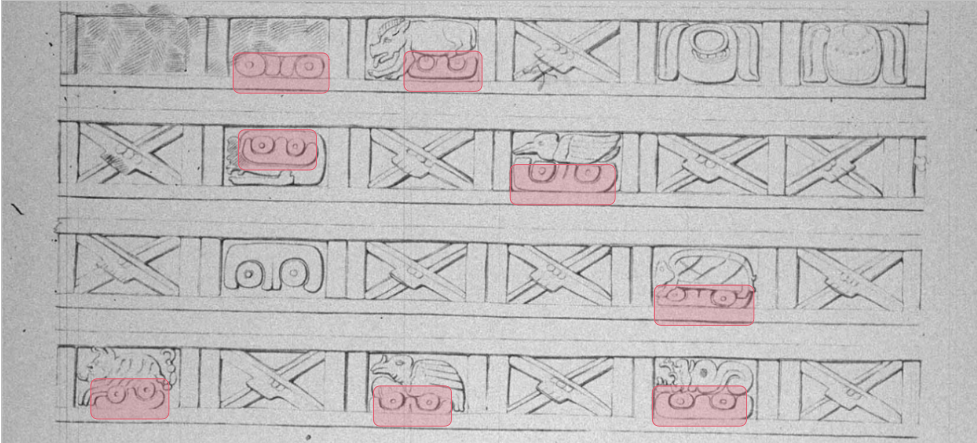}
    \caption{Banda celeste en el Ala Este del complejo de Las Monjas en Chichén-Itzá, dibujado por la artista Annie Hunter, con el glifo Ek’ debajo de las figuras animales y remarcadas en color magenta (\citeauthor{amoud1902}, \citeyear{amoud1902}: Lámina 13).}
\label{Art004:f6}
\end{figure}

Otro ejemplo de representaciones celestes en monumentos lo tenemos en el edificio conocido como Las Monjas, en Chichén Itzá.  En el ala Este, están talladas varias franjas de figuras y glifos, similares a las bandas celestes en los códices.  Algunas de estas figuras son representaciones de animales, similares a los que se muestran en el Códice de París y ordenados de la misma forma que en este documento (\citeauthor{aveni}, \citeyear{aveni}: 278). Debajo de estas figuras zoomorfas también se aprecia el glifo Ek’, asociado al planeta Venus en contextos como la Tabla de Venus en el Códice de Dresde, pero también relacionado al concepto de estrella en otros, lo que refuerza la idea de que estas bandas y las figuras contenidas en ellas tienen un carácter celestial (Figura \ref{Art004:f6}).  Se concluye que son constelaciones en el cielo y que tienen un patrón, observado tanto en las inscripciones de Las Monjas como en el Códice de París.  En el primero se refleja su relación con el movimiento aparente de Venus, porque van acompañando a este planeta que, desde nuestra perspectiva terrestre, aparece como una estrella brillante en el cielo (\citeauthor{aveni}, \citeyear{aveni}: 279).  En el segundo se refleja su relación con el Sol, debido a que todas estas figuras están asociadas al glifo \textit{k’in} o Sol en dicho documento.

\subsection{\textsl{Programa de cómputo \textit{Stellarium}}}

A partir de fines del siglo XX, gracias a los avances en los sistemas informáticos, ha sido posible simular la posición de los cuerpos astronómicos en la esfera celeste, vistos desde cualquier lugar de la superficie terrestre, en escalas de tiempo que van del pasado remoto (según la escala humana) hasta muchos años en el futuro.  Uno de los programas de cómputo de más uso en la actualidad es Stellarium, un proyecto que surge a principios del siglo XXI impulsado por el francés Fabien Chéreau como un proyecto de código abierto.  A lo largo de los años fueron incorporando más colaboradores y hasta el año 2019 el principal responsable de mantenimiento es Alexander Wolf, con la colaboración de Georg Zotti, quienes han implementado nuevas funcionalidades para uso en estudios de Astronomía Cultural. Como parte del uso de Stellarium para estudios de culturas antiguas, se ha incorporado un módulo de “Leyenda Estelar”, la cual permite, a quien esté interesado, incorporar nuevas constelaciones visibles en la esfera celeste, según la cultura que el usuario desee.  Hemos aprovechado esta capacidad para incorporar la civilización maya a la base de datos de culturas, realizando el procedimiento definido en la Guía de Usuario de Stellarium \citep{zotti2019}, así como a través de comunicaciones directas con los actuales responsables del mantenimiento de este programa de cómputo.  Este procedimiento se puede resumir de la siguiente forma:

\begin{enumerate}
    \item Adaptación de imágenes de las constelaciones a formato .png, definiendo un fondo negro en las imágenes y los trazos en color blanco, utilizando un programa de manipulación de imágenes,
    \item Creación de archivos para: nombres de constelaciones, nombres de estrellas brillantes, estrellas en las cuales se fijarían las imágenes de constelaciones y para definir las líneas de las constelaciones,
    \item Creación de archivo de descripción general de la civilización maya y sus tradiciones relativas al espacio celeste,
    \item Incorporar la información mencionada en los incisos anteriores al repositorio oficial de Stellarium, el cual fue revisado por los responsables de mantenimiento, quienes hacían observaciones y mejoras a implementar,
    \item Aprobación e incorporación en la última versión de la información relativa a la cultura maya.
\end{enumerate}

\section{\textsl{Análisis}}

La información recopilada sobre constelaciones mayas tuvo que ser interpretada por los autores, no solo en cuanto a lo que podrían haber significado para los antiguos mayas, sino que también para comprender claramente el significado que varios mayistas quisieron dar a sus lecturas de la información disponible.  Sólo de esta manera se podrían ubicar en el cielo las figuras de las constelaciones sobre las estrellas que más probablemente habrían servido a los antiguos mayas para imaginarlas.  En cuanto a la interpretación de las imágenes en el Códice de París, se utilizó como base el trabajo de \cite{spotak}, ya que existe todavía mucha discusión sobre qué animales o seres representan las figuras en el documento antes mencionado y este autor da explicaciones muy detalladas para cada imagen sobre sus posibles significados.  Estas explicaciones se dan en la Tabla \ref{Art004:tabla1}.  Linda Schele provee las posiciones generales de cada figura en el cielo y las relaciona con las constelaciones adoptadas por la Unión Astronómica Internacional (UAI) (\citeauthor{freidel}, \citeyear{freidel}), pero para ubicar exactamente las figuras en el cielo fue necesario realizar una interpretación propia por los autores.  Se decidió colocar cada figura en el cielo de modo que:

\begin{enumerate}

    \item Su tamaño fuera similar a la de las modernas constelaciones. Debido a que la cantidad de las figuras imaginadas por los mayas (13) junto a la trayectoria aparente del Sol en el cielo (eclíptica) es muy similar a la de las constelaciones zodiacales de la UAI (12), se deduce que dichas figuras deberían tener un tamaño similar para distribuirse uniformemente por la eclíptica.  En ese sentido, se insertaron las figuras para que guardaran una relación de proporcionalidad entre su ancho y su largo similar a como aparecen el Códice de París.
    \item Ya que en este códice se observa que las bocas o fauces de los seres representados están justo por debajo del glifo \textit{k’in}, se les decidió colocar en una ubicación que respetara el paso del Sol por la posición de las bocas o fauces de cada figura, es decir, la eclíptica siempre está por sobre las figuras de las constelaciones, a lo largo de todo el cielo.  Ya que dichas constelaciones están dispuestas siguiendo la posición del Sol y son en su mayoría zoomorfos, cabe llamarlas en su conjunto como “zodíaco maya”.

\end{enumerate}

\begin{table}
    \centering
        \caption{\centering Descripción de las constelaciones incorporadas al cielo virtual del programa de cómputo Stellarium a partir del Códice de París \citep{spotak}.}
        \label{Art004:tabla1}
        
        \begin{tabular}[t]{cccl}
        \toprule

        \textbf{Imagen} & \textbf{Yucateco} & \textbf{Español} & \textbf{Descripción}\\

        \hline \\ [-1.5ex]

        \includegraphics[width=0.1\textwidth, height=20mm]{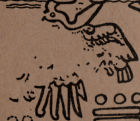}
         & \small Ch’oom
         & \small Zopilote
         & \begin{minipage}{0.6\textwidth}
             \small Esta figura no ha sido claramente identificada, \citeauthor{tozzer1910}
             (\citeyear{tozzer1910}:  330-331) propone la conexión mitológica
             con el zopilote, ya que no está decorada y se relaciona
             muy bien con fuentes sobre este animal en el Chilam
             Balam of Chumayel (\citeauthor{roys1967}, \citeyear{roys1967}: 73).\\ [-1ex]
            \end{minipage}\\
        \hline \\ [-1.5ex]

        \includegraphics[width=0.1\textwidth, height=20mm]{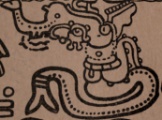}
        & \small Xoc
        & \small Tiburón
        & \begin{minipage}{0.6\textwidth}
          \small Esta imagen es en realidad la combinación de los elementos de dos animales: la cabeza de una serpiente y la cola de un pez. Schele lo identifica con la representación de un tiburón (\citeauthor{villelas1993}, \citeyear{villelas1993}: 38), pudiendose identificar ambas características en este animal. \\ [-1ex]
        \end{minipage} \\
        \hline \\ [-1.5ex]

        \includegraphics[width=0.1\textwidth, height=20mm]{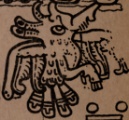}
         & \small Kulte’
         & \small Búho
         & \begin{minipage}{0.6\textwidth}
           \small Milbrath, Schele, Love y Thompson, lo consideran un búho.  En particular lo identifican con el \textit{Bubo} \textit{virginianus} (Búho cornudo) o el \textit{Otus guatemaleae}       (Autillo guatemalteco) (\citeauthor{bb2011}, \citeyear{bb2011}: 699). Esto viene apoyado por las similares imágenes de pájaros que aparecen en las páginas k'atun. \\ [-1.5ex]
         \end{minipage} \\
        \hline \\ [-1.5ex]

       \includegraphics[width=0.1\textwidth, height=20mm]{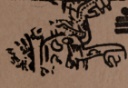}
        & \small Balaam
        & \small Jaguar
        & \begin{minipage}{0.6\textwidth}
        \small Identificado como tal debido a las típicas manchas que exhibe este animal sobre su piel, así como a las garras y la cola que son claramente visibles en esta imagen del Códice de París.\\ [-1ex]
        \end{minipage} \\
        \hline \\ [-1.5ex]

        \includegraphics[width=0.1\textwidth, height=20mm]{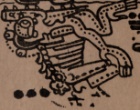}
         & \small Kimi
         & \small Muerte
         & \begin{minipage}{0.6\textwidth}
          \small Aparece claramente como un esqueleto, sin embargo, no está claro si pertenece a un animal o a un humano, posiblemente se refiere al concepto general de muerte. \\ [1ex]
          \end{minipage} \\
        \hline \\ [-1.5ex]

       \includegraphics[width=0.1\textwidth, height=20mm]{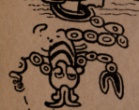}
        & \small Sina’an
        & \small Escorpión
        & \begin{minipage}{0.6\textwidth}
        \small En este caso aparece claramente la imagen de un escorpión el que es muy frecuente en la iconografía maya. El códice de Madrid y las cerámicas K1126 y K4525 de la base de datos de cerámica de Justin Kerr son solamente algunos ejemplos. En el cielo, este animal está en el sentido opuesto a como se visualiza en la cultura occidental, es decir, la estrella Antares está en el aguijón de la constelación maya.
        \end{minipage} \\
        \hline \\ [-1.5ex]

        \includegraphics[width=0.1\textwidth, height=20mm]{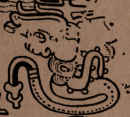}
         & \small Chaan
         & \small Serpiente
         & \begin{minipage}{0.6\textwidth}
         \small Este animal se encuentra presente en América Central en forma de varias especies, pero el mostrado en el Códice de París corresponde a una cascabel, como se aprecia por el crótalo de la figura.\\
         \end{minipage} \\
        \hline \\ [-1.5ex]

        \includegraphics[width=0.1\textwidth, height=20mm]{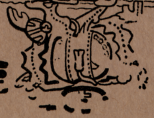}
         & \small Zotz
         & \small Murciélago
         & \begin{minipage}{0.6\textwidth}
         \small Apoyado por la presencia del glifo T523, la forma de las alas y por la característica forma de la nariz que se observa en la figura, muy típica en la iconografía maya de este animal.\\
         \end{minipage} \\
        \hline
        \end{tabular}
\end{table}

Respecto a la ubicación de constelaciones que no estuvieran alineadas siguiendo el patrón de la eclíptica, se utilizó nuevamente como referencia el trabajo de Linda Schele, quien en (\citeauthor{freidel}, \citeyear{freidel}) da un recuento sobre cómo dedujo la relación existente entre la Vía Láctea y las constelaciones de Orión y Géminis con la cosmogonía maya, específicamente, en cuanto al papel que jugaron en ésta el cocodrilo-venado-estelar, la tortuga cósmica, el fuego primordial de la hoguera de las tres piedras y el dios del maíz en las historias de la creación.  \citeauthor{galindo2009} (\citeyear{galindo2009}: 89-90) provee más evidencias que apoyan esta interpretación al analizar uno de los murales del Cuarto II del Edificio de las Pinturas en Bonampak, México (ver Figura \ref{Art004:f4}). Este investigador relaciona la orientación del Edificio con la del cielo en la fecha a la que corresponde el evento de juzgamiento llevado a cabo por el gobernante en el mural, como se deduce de los glifos que están alrededor de esta escena, según los cuales, ocurrió el 6 de agosto del 792 d.C.  Siempre según este autor, al atardecer y amanecer de este día, la Vía Láctea estaba alineada con el eje principal de esta estructura, lo que explica la figura del monstruo celeste que se aprecia en forma de banda celeste en las cerraduras de cada cuarto pintado de este edificio.  La parte visible de la Vía Láctea en el amanecer de esa fecha es la que corresponde con el brazo de Orión, donde están las constelaciones de Orión y Géminis, con los planetas Saturno y Marte en medio de ambas. Esta disposición coincide con el de las figuras dibujadas en el mural antes mencionado, que está a su vez, orientado en la misma forma que nuestra galaxia como se veía en esa fecha, por lo que se pueden asociar las figuras con los correspondientes astros en el cielo en ese momento teniendo a la Vía Láctea como fondo.  Esto permite ubicar las figuras de los pecaríes y de la tortuga en las constelaciones de Géminis y Orión, respectivamente.  En el caso de la tortuga es más sencillo, ya que simplemente se hacen coincidir las estrellas sobre su caparazón con las estrellas del cinturón de Orión, mientras que los pecaríes se hacen coincidir ambos entre las estrellas de Pollux y Castor. Sin embargo, y como ocurre en la mayoría de las ocasiones, lastimosamente no poseemos analogías que permitan falsar estas hipótesis. Para una mayor discusión sobre la orientación del edificio y la interpretación realizada sobre este edificio conviene dirigirse al trabajo de \citeauthor{sanchezs2015} (\citeyear{sanchezs2015}: 35) que obtuvieron para las coordenadas geográficas N16.7036° y W91 0651°, un acimut de 35.091° y una declinación de 54.229°.

\begin{figure}[p]
    \centering
    \includegraphics[scale=0.4]{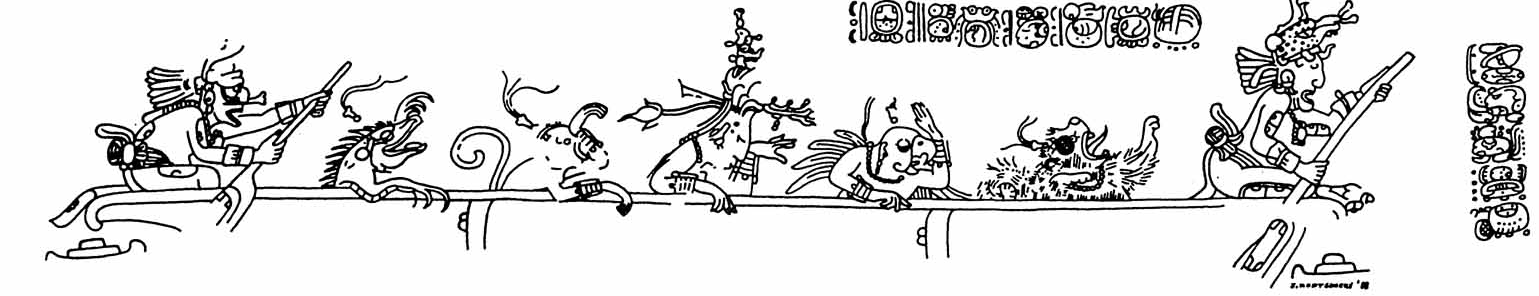}
    \caption{Hueso JM00780 del Entierro 116 de Tikal. Dioses remeros llevando al dios del Maíz hacia su sacrificio en el fuego primordial. Dibujo de John Montgomery.}
\label{Art004:f7}
\end{figure}

Retomando el tema de la cosmogonía maya, la tortuga es la que se relaciona con este acontecimiento de creación.  Según representaciones de pasajes del Popol Vuh en vasijas mayas, el dios del maíz surge del caparazón de la tortuga, renacido gracias al esfuerzo de sus dos hijos, Hunahpú e Ixbalanqué. Por tanto, alrededor de esta constelación debería ocurrir la historia de creación, ubicando el fuego primordial al par de la tortuga, ya que las tres estrellas de la tortuga representan también, en el imaginario maya, las tres piedras del hogar, que en esta cultura se utilizan para hacer la fogata central en las viviendas. Según las historias de creación, el dios del maíz tuvo previamente que ofrecerse en sacrificio llevado en una canoa por un par de dioses remeros y acompañados de animales que lamentan la pronta muerte de este dios.  Esta historia se representa en el cielo, según Schele (\citeauthor{freidel}, \citeyear{freidel}) de la siguiente forma: el fuego primordial es la nebulosa de Orión (M42), ya que a simple vista y desde lugares muy oscuros (como debieron ser los cielos que tuvieron los antiguos mayas) se le observa como una pequeña mancha difusa, similar al humo que sale de una fogata.  Favorece a esta idea el hecho que esta nebulosa está muy cerca del cinturón de Orión, por lo que se forma el escenario perfecto para la ubicación de la fogata con sus tres piedras.  Los dioses remeros transportan al dios del maíz a través de la Vía Láctea, en dirección hacia el fuego primordial, donde será su sacrificio, por tanto, se colocó la imagen de los dioses remeros transportando al dios del maíz (Figura \ref{Art004:f7}).

\begin{figure}[p]
    \centering
    \includegraphics[scale=0.85]{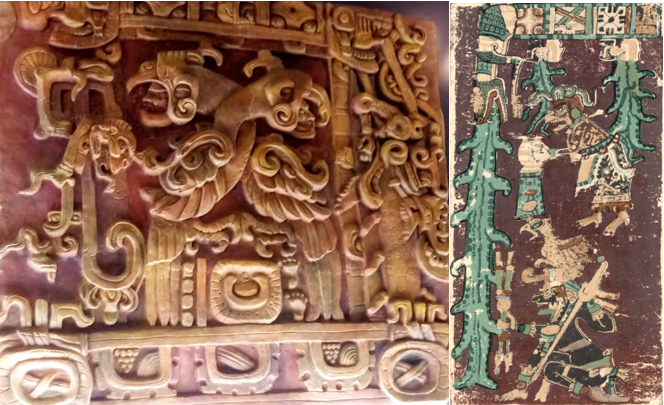}
    \caption{Fachada de templo Margarita en Copán (reproducción en el Museo de la Identidad Nacional, Honduras) y página 74 del Códice de Dresde, ambos mostrando el monstruo cósmico colgado de la banda celestial y derramando agua sobre el mundo. Fotografía izquierda por el autor, fotografía derecha, tomada de \citep{FAMSIb}}
\label{Art004:f8}
\end{figure}

La última constelación que se trabajó fue la que relaciona a la Vía Láctea con el monstruo primordial llamado el cocodrilo-venado-constelado o cocodrilo-venado-estelar.  Este ser, en otra versión de la creación en la mitología maya, fue el causante de un diluvio sobre la tierra, al hacer brotar una gran cantidad de agua de sus fauces.  El dios GI (todavía por descifrar su verdadera identidad y significado) atrapó a este monstruo, lo decapitó y con su sangre, creó al mundo.  Este evento fue tan importante en el imaginario maya, que los gobernantes lo utilizaron en la iconografía que encargaban para conmemorar su ascenso al poder, justificando dicho poder al relacionarlo con la capacidad de dominar los elementos (representados por el cocodrilo-venado-estelar) tal como hicieron los dioses GI, GIII o Chaak, algo que está todavía por definir por parte de los investigadores (\citeauthor{garciab2015}, \citeyear{garciab2015}: 20).  Existen varias representaciones de este monstruo cósmico, tales como en la fachada del templo Margarita en Copán y en el Códice de Dresde (Figura \ref{Art004:f8}). Se seleccionó la misma que usa Linda Schele en su explicación de cómo ella llegó a la conclusión que según la época del año, la Vía Láctea es una representación de este ser, que corresponde al monstruo de la Estela 25 del sitio de Izapa (Figura \ref{Art004:f5}) (\citeauthor{freidel}, \citeyear{freidel}: 88-89).

\section{Resultados}

\begin{figure} [p]
    \centering
    \includegraphics[scale=0.8]{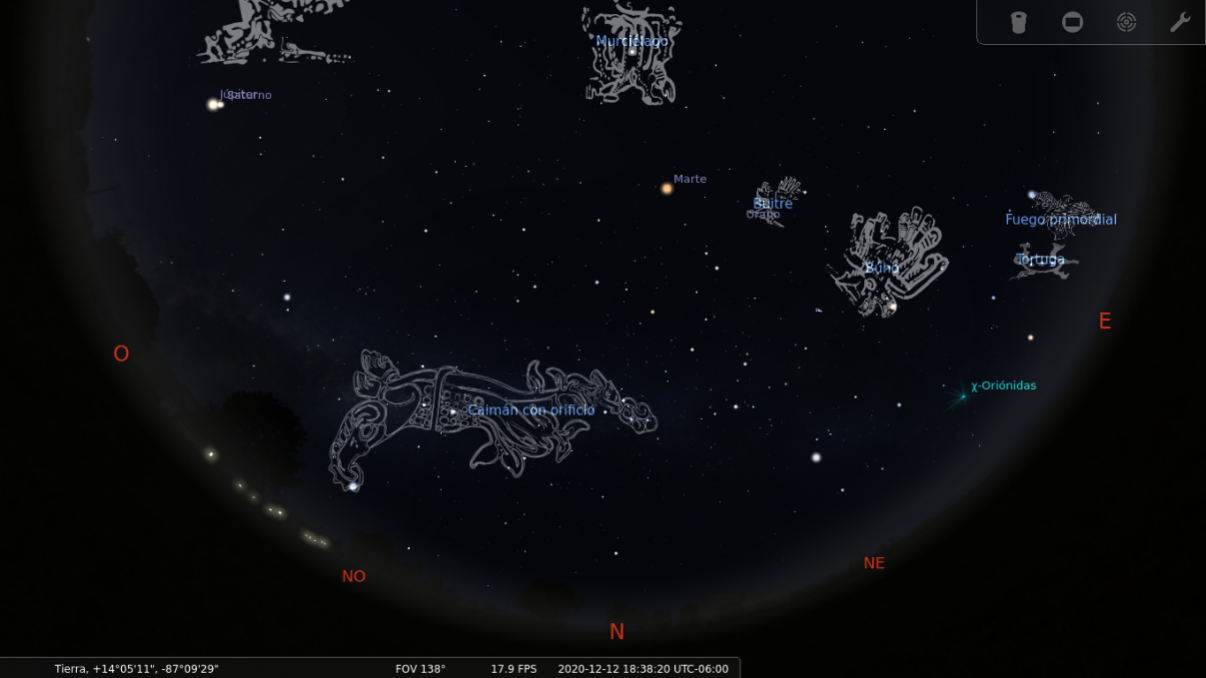}
    \caption{Constelaciones mayas observables al Norte cerca del Solsticio de Invierno, poco después del ocaso.}
\label{Art004:f9}
\end{figure}

\begin{figure} [p]
    \centering
    \includegraphics[scale=0.75]{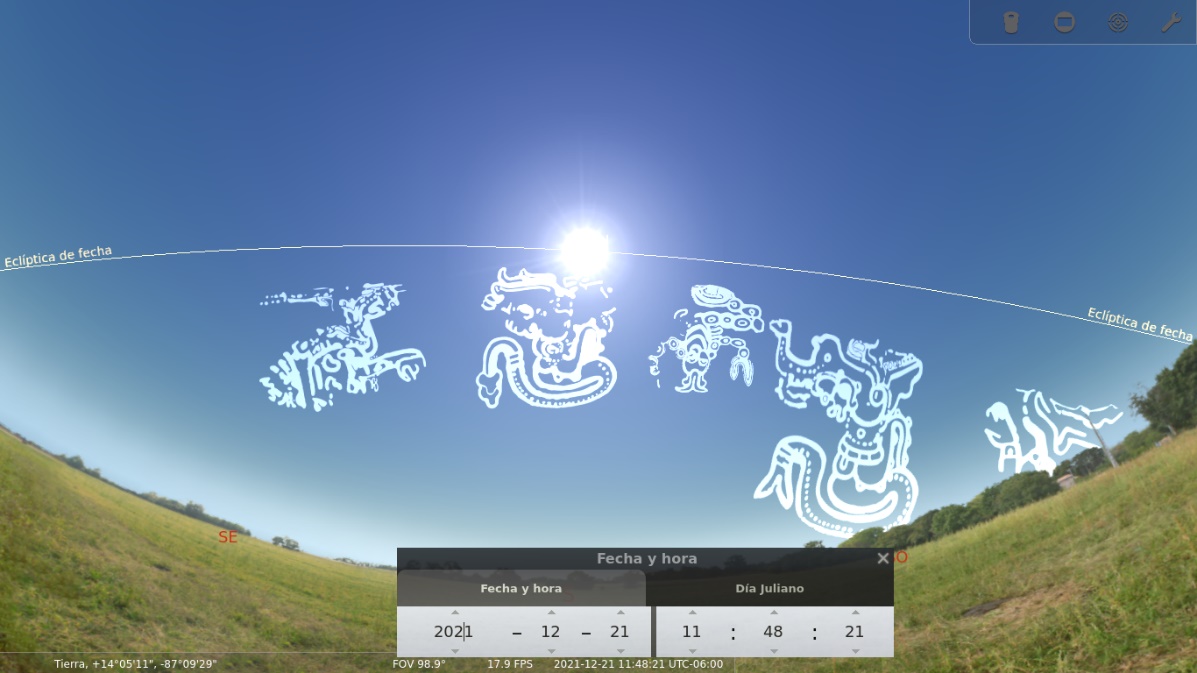}
    \caption{Constelaciones mayas observables hacia el Sur en el Solsticio de Invierno, hacia el mediodía, mostrando la eclíptica (ruta aparente del Sol sobre el cielo) y su posición relativa respecto de las constelaciones mayas.}
\label{Art004:f10}
\end{figure}

\begin{figure} [p]
    \centering
    \includegraphics[scale=0.75]{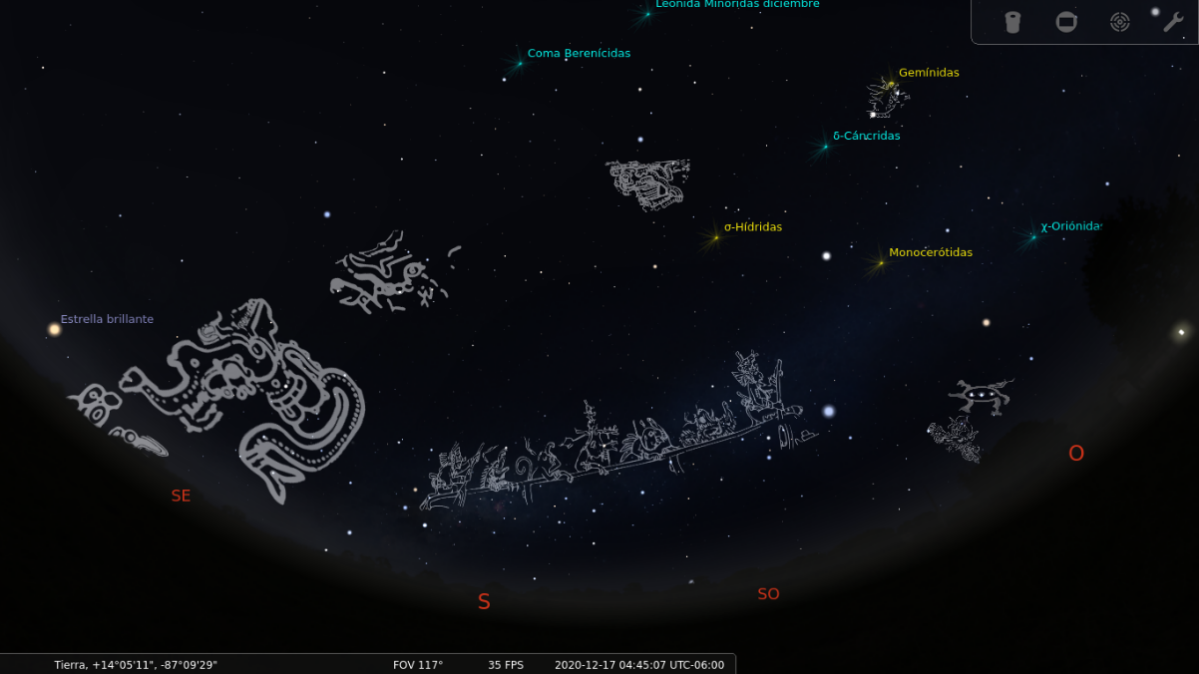}
    \caption{Constelaciones representando la Cosmogonía Maya, observables poco antes del amanecer, cerca del Solsticio de Invierno.}
\label{Art004:f11}
\end{figure}

\begin{figure} [p]
    \centering
    \includegraphics[scale=0.75]{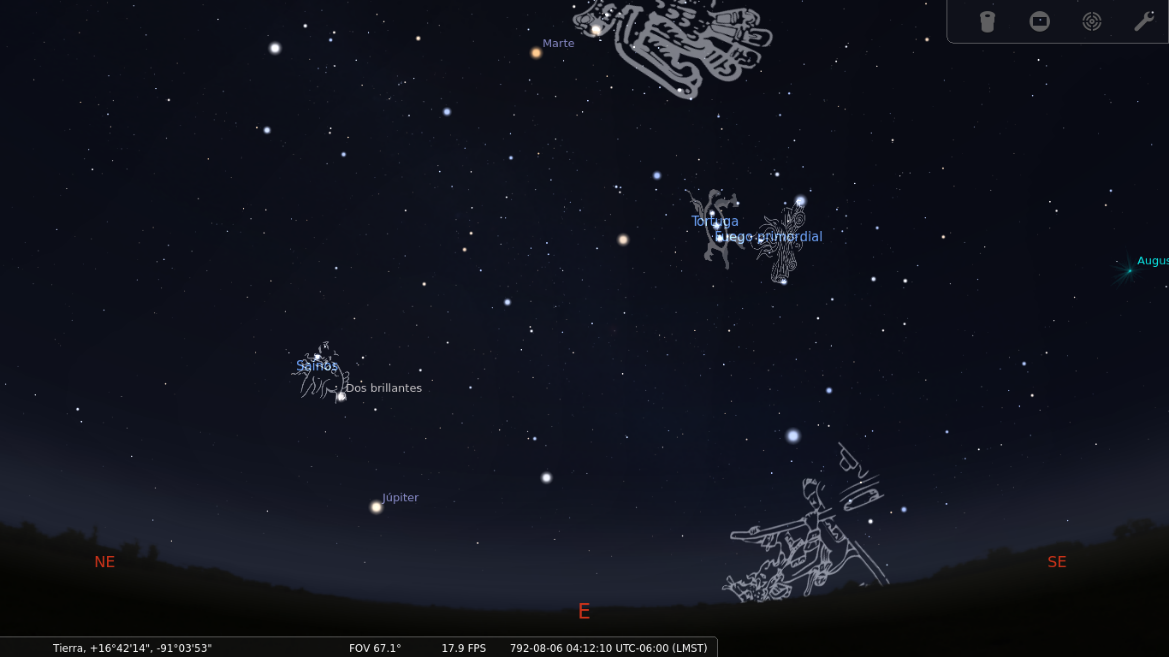}
    \caption{Constelaciones visibles hacia el Este, al amanecer del 6 de agosto del 792 d.C., en la ciudad maya de Bonampak.}
\label{Art004:f12}
\end{figure}

Producto del trabajo descrito en las secciones anteriores, se obtuvo una representación del cielo que consideramos es cercana a lo que vieron e imaginaron los antiguos mayas. Tras la recolección de la información recogida de las diferentes fuentes explicadas en los epígrafes superiores, se procedió a la implementación de las constelaciones mayas en el programa Stellarium. Para la comprobación del funcionamiento óptimo de esta nueva característica del programa, se representó el cielo para diferentes épocas del año, logrando obtener paisajes que proponemos son similares a lo que los antiguos mayas observaron en sus cielos (Ver Figuras \ref{Art004:f9}, \ref{Art004:f10}, \ref{Art004:f11} y \ref{Art004:f12}).  El punto de observación es Tegucigalpa, Honduras (Latitud 14° 05’ 11” N, Longitud 87° 09’ 29” W), para las figuras \ref{Art004:f9}, \ref{Art004:f10} y \ref{Art004:f11}.  La figura \ref{Art004:f12} corresponde a las constelaciones como se observaron el día 6 de Agosto del año 792 d.C. desde la antigua ciudad maya de Bonampak (Latitud 16° 42’ 14” N, Longitud 91° 03’ 53” W),  fecha que corresponde con la inscrita en los famosos murales dentro de las bóvedas del edificio de las pinturas. Este último se incluye con el objetivo de compararlas con la narración y la interpretación que de estos murales han realizado diferentes investigadores (\citeauthor{freidel}, \citeyear{freidel}; \citeauthor{galindo2009}, \citeyear{galindo2009}). Se puede apreciar una aparente correlación entre varias constelaciones como los pecaries y la tortuga. Si esto fuera así, los personajes que aparecen en el mural de Bonampak representarían planetas, siendo el primer caso, el planeta Júpiter, que se situaría a la derecha de los pecaríes.

\section{Conclusiones}
Este trabajo ha recolectado un gran número de fuentes sobre las constelaciones mayas en una representación gráfica, aunque obviamente esta puede no ser única ni completa. Por otro lado, las fuentes que han llegado hasta nuestros días no cubren el vasto espectro cultural de los pueblos mayas, pero sí permite tener una idea más cercana a cómo plasmaban su cosmovisión en el espacio celeste. Este tipo de herramientas se instituyen como una forma de explorar las posibles constelaciones que se identificaron como parte del discurso cosmogónico para la cultura maya.

Además del uso académico de estos datos, este trabajo tiene un componente claramente didáctico a la hora de introducir tanto la mitología, la cosmogonía, la cosmovisión y conceptos astronómicos de los pueblos tantos mayas como mesoamericanos.

\bibliographystyle{unsrtnat}
\bibliography{biblioArt004}  


\end{document}